\documentstyle[sprocl]{article}

\def\be{\begin{equation}}
\def\ee{\end{equation}}
\def\bea{\begin{eqnarray}}
\def\eea{\end{eqnarray}}

\begin{document}

\title{Renormalization of the meson-baryon SU(3) Lagrangian in
  chiral perturbation theory to order q$^3$}

\author{GUIDO M\"ULLER \footnote{e-mail:
    mueller@pythia.itkp.uni-bonn.de}}

\address{Institut f\"ur Theoretische Kernphysik \\
Nussallee 14 - 16 \\
53115 Bonn, Germany}




\maketitle\abstracts{
We perform the complete regularization of
all Green functions with a single incoming and outgoing baryon to order
q$^3$ in the low energy expansion 
for the SU(3) meson-baryon system in chiral perturbation theory.
The divergences can be
extracted in a chiral invariant manner by making use of the heat
kernel representation of the propagators in d-dimensional Euclidean
space.
}

In the arena of low energy physics it is much more useful to introduce
an effective Lagrangian, which is written in terms of experimental
degrees of freedom --mesons and baryons-- and which encodes the
symmetries of the underlying QCD interaction -- specifically for our purpose
chiral symmetry, which exists in the limit in which the quark mass can
be taken as vanishing.
Since we are interested in processes were the momenta are small, we
can expand the Green functions in powers of the external fields. 
The low energy expansion
involves two small parameters, the external momenta $q$ and the quark
masses ${\cal M}$. One expands in powers of these with the ratio
${\cal M}/q^2$ fixed. 
The low energy expansion is now obtained from a perturbative expansion
of the meson EFT,
\be
{\cal L}_M = {\cal L}_M^{(2)}  + {\cal L}_M^{(4)} + \dots
\label{effexpansion}
\ee
where the index ($ d = 2,4,\dots $) denotes the low energy
dimension. The leading term in the low energy expansion
(\ref{effexpansion}) can be easily written down in terms of the
pseudoscalar Goldstone fields ($\phi = \pi, K, \eta$)  
which are collected in
a $3 \times 3$ unimodular, unitary matrix $U(x)$, 
\begin{equation}
 U(\phi) = u^2 (\phi) = \exp \lbrace i \phi  / F_0 \rbrace
\label{U}
\end{equation}
transforming linearly under chiral symmetry.
The lowest order Lagrangian consistent with Lorentz
invariance, chiral symmetry, parity, G-parity and charge conjugation
reads 
\be
{\cal L}^{(2)}_M = \frac{F^2_0}{4} ( < \partial_\mu U  \partial^\mu
U^\dagger  > + 2 B_0 < 
{\cal M} ( U + 
U^\dagger) > )
\label{ltwo}
\ee
The scale $F_0$ can be identified with the
pseudoscalar decay constant in the chiral limit and the
constant $B_0$ is related to the
quark condensate. 
One can now calculate tree diagrams
using the effective Lagrangian ${\cal L}_M$ and derive all 
current algebra predictions. One has to include higher order
corrections and meson loops, because tree diagrams are always real and
thus unitarity is violated. Weinberg made the important observation
that diagrams with $ L \, (L=1,2,\dots) $ meson loops are suppressed by
powers of $(q^2)^L$ with respect to the leading term.
In the meson
sector we have a one--to--one correspondence between the number $L$ of
loops and the chiral dimension
\be
D= 2 + \sum_d N_d(d-2) + 2 L \, 
\label{mesoncounting}
\ee
where $N_d$ denotes the number of vertices of dimension $d$.
For a given $S$-matrix element, the chiral dimension $D$ increases
with the number of meson loops. 
It should be not surprising that a generic one-loop diagram in CHPT is
divergent. The divergences can be taken care of once and for all by
calculating the divergent part of the one-loop functional. Since it
must be a local action with the symmetries of $ {\cal L}_M^{(2)}$  and
since it has chiral dimension $D=4$, the corresponding Lagrangian must
be of the general form as discussed in ref \cite{gl84}. 
\be
{\cal L}_4  = \sum_{i=1}^{10} L_i P_i + \sum_{i=1}^{2} H_i
\tilde{P}_i 
\label{leff4}
\ee
The renormalization procedure at
one-loop level is then achieved by simply dividing the coefficients in
(\ref{leff4}) into a scale dependent finite and divergent part:
\be
L_i = L_i^r(\mu) + \Gamma_i \Lambda \, .
\label{deco}
\ee
The constants $\Gamma_i$ serve to renormalize  the infinities of the
meson loops  and the remaining finite pieces $L_i^r(\mu)$ have to be fixed
phenomenologically or to be estimated by some model.    
With dimensional regularization, the divergent factor $\Lambda$ is
given by
\be
\Lambda = \frac{\mu^{d-4}}{(4\pi)^2} \Big\{ \frac{1}{d-4} -
\frac{1}{2} [\log(4\pi) + 1 + \Gamma'(1)] \Big\}
\ee
depending on the arbitrary renormalization scale $\mu$. 
The sum of one-loop amplitudes
and of tree amplitudes based on (\ref{leff4}) is not only finite, but
also independent of the scale $\mu$.  
To evaluate these loop graphs
one considers the neighbourhood of the solution 
to the classical equations of motion. 
The contribution to the one-loop graphs to the generating functional is
given by 
\bea
\exp \{ i {\cal Z}_{1-loop} \} & = &
N  [\det (dd+\sigma)]^{-\frac{1}{2}}  
\label{zoneloop}
\eea
where the explicit definitions are given in \cite{gl84}.
The determinant is UV divergent and
produces poles at $d=0,2,4,\dots$ in dimensional regularization
and can be regularized via the heat kernel expansion.


Let us now consider the meson-baryon interaction in CHPT.
Because of the different Lorentz structure of meson and baryon
fields, the chiral expansion of ${\cal L}_{MB}$ contains terms of
$O(q^n)$ for each positive integer $n$, unlike in the case of ${\cal
  L}_M$ where the chiral expansion proceeds in steps of two  powers of
$q$. So we have
\be
{\cal L}_{MB} =  {\cal L}^{(1)}_{MB} + {\cal L}^{(2)}_{MB} +  {\cal
  L}^{(3)}_{MB} 
+ \dots  \, .
\ee
The effective meson--baryon Lagrangian for processes with
one incoming and one outgoing baryon is bilinear in the baryon field
and starts with terms of dimension one,
\be
{\cal L}_{MB}^{(1)} = <\bar B \, (i \nabla\!\!\!\!/ - m) \, B> 
+ \frac{D}{2} \, < \bar B \, \{u\!\!\!/ \gamma_5 , B \}\,>
+ \frac{F}{2} \, < \bar B \, [ u\!\!\!/ \gamma_5 , B ] \,> \,\,\, .
\label{LMB1}
\ee
We collect the spin--$\frac{1}{2}$ baryon octet
in the $3\times 3$ matrix $B$. 
The lowest order
meson--baryon Lagrangian contains two axial--vector coupling
constants, denoted by $D$ and $F$ and the average baryon octet mass
$m$ in the chiral limit. In contrast,
baryon propagators introduce the baryon mass as an additional
dimensional quantity, which remains finite in the chiral
limit. Consequently, we have not a correspondence between loop
expansion and chiral dimension as in the meson sector expressed in
eq.(\ref{mesoncounting}). With inspiration from heavy quark
effective theory, Jenkins and Manohar \cite{Je91}
have reformulated baryon CHPT in
precisely such a way as to transfer the baryon mass from the
propagator to the vertices by defining velocity-dependent fields.
The chiral dimension $D$ for processes
with exactly one baryon line running through the pertinent Feynman
diagrams is given by \cite{We}
\be
D = 2L +1 + \sum_{d=2,4,6,\ldots} (d-2) \, N_d^M + \sum_{d=1,2,3,\ldots} 
(d-1) \, N_d^{MB} \ge 2L+1
\label{chdim}
\ee
with $N_d^M$ ($N_d^{MB}$) 
counts the number of mesonic (meson--baryon) vertices of dimension $d$.
In contrast to the relativistic formulation, chiral power counting in
the heavy baryon formalism is completely analogous to the mesonic
case. As for the lowest-order constants $F_0,B_0$ in (\ref{ltwo}), the
coupling constants in ${\cal L}^{(1)}_{MB} + {\cal L}^{(2)}_{MB}$ are
not renormalized in any order of the loop expansion, since $D \geq 3
$.
As in the meson-baryon sector we expand around the classical solution of
motion and the  
corresponding one-loop generating functional 
consists of four pieces corresponding to two irreducible and two
reducible diagrams \cite{Ec94} \cite{gss} \cite{mu}. 
It can be shown that the sum of the reducible diagrams is finite and
scale independent. The divergences of the one-loop graphs are therefore
contained in the irreducible diagrams which are given by \cite{Ec94}
\cite{mu}  
\bea
Z_{\rm irr}[j,R_v^a] &=& \int d^4x \, d^4x' \, d^4y \, d^4y' \, 
\bar{R}_v^a (x) \, S_{(1)}^{bc, \,\rm cl} (x,y) \times \nonumber\\
& & \!\! \bigl[ \, \Sigma_2^{cd}(y,y) \,  \delta(y-y')
+ \Sigma_1^{cd} (y,y') \, \bigr] \, S_{(1)}^{de, \, \rm cl}(y',x')
\, R_v^e(x')
\label{Zirr}
\eea
$ \Sigma_1 $ corresponds to self-energy diagrams and $ \Sigma_2 $ contains
tadpol diagrams. Note that the propagators and the vertices have the
full tree-level structure attached to them as functionals of the
external fields.
The regularization in the meson-baryon
sector is extremely tedious, because we are not able to express the
one-loop functional in terms of a determinant unlike the meson sector.  
Nevertheless the divergences can be
extracted in a chiral invariant manner by making use of the heat
kernel representation of the propagators in d-dimensional Euclidean
space. In the case of the self--energy contribution the divergences
are due to the singular behaviour of the product of the meson and the
baryon propagators in the conincidence limit. 
A non-trivial part of the regularization consists in finding a heat
kernel representation for the full tree-level baryon propagator \cite{Ec94}.
The main difference between the SU(2) 
 and the SU(3) calculation lies in the fact that the nucleons are in
the fundamental representation of SU(2), while the baryons are in the
adjoint representation of SU(3). This group theoretical difference in
flavor space does not change the singular behaviour in coordinate
space  which leads to the divergences.  
The complete renormalization of Green functions for
the pion-nucleon SU(2) interaction was done by Ecker \cite{Ec94} 
and we performed the regularization in SU(3) CHPT\, \cite{mu}. 
To separate the finite parts in dimensional regularization, we follow
the conventions of \cite{gl85} to decompose the irreducible one--loop
functional into a finite and a divergent part. Both depend on the
scale $\mu$.
The generating functional can then be renormalized by introducing the
counterterm Lagrangian \cite{mu}
\be \label{LCT}
{\cal L}_{MB}^{(3)\, {\rm ct}} (x) = \frac{1}{(4 \pi F_0 )^2} \,
\sum_{i=1}^{102} 
d_i \, \bar{H}^{ab}_v (x) \, \tilde{O}^{bc}_i (x) \, H^{ca}_v (x)
\ee
where the $d_i$ are dimensionless coupling constants and the field
monomials $\tilde{O}^{bc}_i (x)$ are of order $q^3$. The low--energy 
constants $d_i$ are decomposed in analogy to Eq.(\ref{deco}).
The $\beta_i$ are dimensionless functions of $F$ and $D$ constructed
such that they cancel the divergences of the one--loop functional. 
The counterterms constitute a complete set for the 
renormalization of the irreducible tadpole and self--energy functional
for {\it off-shell} baryons. These are the terms where the covariant 
derivative acts on the baryon fields. As long as one is only
interested in Green functions with {\it on-shell} baryons, the number
of terms can be reduced considerably by invoking the baryon equation
of motions. 
A further reduction in
the number of terms could be achieved by use of the Cayley--Hamilton 
relation. Also, many of the terms given in the
table refer to processes with at least three Goldstone bosons. These
are only relevant in multiple pion or kaon production by photons or
pions off nucleons \cite{bkmpipin}.
The renormalized LECs $d_i^r (\mu)$ are measurable quantities. They
satisfy the renormalization group equation.
Therefore, the choice of another scale $\mu_0$ leads to modified
values of the renormalized LECs.
We remark that the scale--dependence in the counterterm Lagrangian is,
of course, balanced by the scale--dependence of the renormalized
finite one--loop funcional for observable quantities.

This method allows
to control the full divergence structure of $O(q^3)$ and permits a
complete renormalization of {\bf all} Green functions: baryon form
factors, photo(electro) kaon production, kaon-baryon scattering, etc.
In summary, we remark that the method used has the disadvantage of
leading to very lengthy and complicated expressions in the
intermediate steps due to the loss of covariance. There should exist
an improved method which does not share this complication. 
In particular, the necessary
inclusion of virtual photons in the pion--nucleon (or meson--baryon)
system can be treated along these lines. We hope to report on the
results of such an investigation in the near future.

\section*{Acknowledgments}
I am grateful to Ulf-G.\,Mei\ss ner for helpful discussions.

\end{document}